# New Results on the Number Theoretic Hilbert Transform

Vamsi Sashank Kotagiri
Oklahoma State University, Stillwater

**Abstract**
This paper presents new results in the theory of number theoretic Hilbert (NHT) transforms. New polymorphic solutions have been found for the 14-point and 16-point transforms. Several transform pairs are computed and solutions found for which the sequence and the transform have the same shape. The multiplicity of solutions for the same moduli increases their applicability to cryptography.

*Keywords:* Hilbert transforms, number theoretic transforms, scrambling, data security

**Introduction**
The discrete Hilbert transform (DHT) [1], is a circulant matrix with alternating entries of each row being zero and non-zero numbers and transpose modulo a suitable number is its inverse, and it has applications is a variety of fields, including signal processing [2]-[6] and cryptography [8]-[10]. An early unsuccessful attempt to create a number-theoretic Hilbert transform (NHT) was given in [11], and a successful structure, with particular potential for cryptography due to its polymorphic solutions, was given in [12], with additional results in [13], where NHT matrices were presented for up to 12-points.

Our notation is as follows: the data block is the vector F, and the NHT transformed data block is the vector G, the NHT transform is the matrix N, and the computations are with respect to the modulus m, the inverse of the NHT matrix is $N^T$ mod m. Thus

$$G = NF \bmod m, \text{ and}$$
$$F = N^T G \bmod m \quad (1)$$

In the present paper we extend the results of the previous paper [12] by providing equations for 14-point and 16-point NHT. Several transform pairs are computed and solutions found for which the sequence and the transform have the same shape.

**14-point NHT**
The first row of the 14-point NHT matrix will be given by integers a, b, c, d, e, f, g that alternate with 0s. The problem is to find the circulant matrix with these values in the first row that satisfies the conditions given by (1). When we multiply the 14-point NHT with its transpose we observe that the squares of the non-zero integer values of the first row should equal 1 modulo the chosen m. In other words,

$$a^2 + b^2 + c^2 + d^2 + e^2 + f^2 + g^2 = 1 \bmod m \tag{2}$$

The other non-diagonal element terms in the product $NN^T$

$$\begin{aligned} &ab + bc + cd + de + ef + fg + ga \\ &ac + bd + ce + df + eg + fa + gb \\ &ad + be + cf + dg + ef + fg + ga \end{aligned} \tag{3}$$

should be all zero with respect to the same modulus.

In order to get a valid NHT matrix we need to assume a suitable modulus in such that all the non-diagonal elements of the matrix product $NN^T$ will become zero and only the diagonal elements of the product matrix remain.

There are many solutions which satisfy the equation if we randomly choose the values of a = 3 b =15 c = 22 d = 11 e =20 f=10 and g=5 and thus we can write the 14-point NHT transformation as G = HT mod 29 which is shown as below:

$$\begin{bmatrix} g(0) \\ g(1) \\ g(2) \\ g(3) \\ g(4) \\ g(5) \\ g(6) \\ g(7) \\ g(8) \\ g(9) \\ g(10) \\ g(11) \\ g(12) \\ g(13) \end{bmatrix} = \begin{bmatrix} 0 & 3 & 0 & 15 & 0 & 22 & 0 & 11 & 0 & 20 & 0 & 10 & 0 & 5 \\ 5 & 0 & 3 & 0 & 15 & 0 & 22 & 0 & 11 & 0 & 20 & 0 & 10 & 0 \\ 0 & 5 & 0 & 3 & 0 & 15 & 0 & 22 & 0 & 11 & 0 & 20 & 0 & 10 \\ 10 & 0 & 5 & 0 & 3 & 0 & 15 & 0 & 22 & 0 & 11 & 0 & 20 & 0 \\ 0 & 10 & 0 & 5 & 0 & 3 & 0 & 15 & 0 & 22 & 0 & 11 & 0 & 20 \\ 20 & 0 & 10 & 0 & 5 & 0 & 3 & 0 & 15 & 0 & 22 & 0 & 11 & 0 \\ 0 & 20 & 0 & 10 & 0 & 5 & 0 & 3 & 0 & 15 & 0 & 22 & 0 & 11 \\ 11 & 0 & 20 & 0 & 10 & 0 & 5 & 0 & 3 & 0 & 15 & 0 & 22 & 0 \\ 0 & 11 & 0 & 20 & 0 & 10 & 0 & 5 & 0 & 3 & 0 & 15 & 0 & 22 \\ 22 & 0 & 11 & 0 & 20 & 0 & 10 & 0 & 5 & 0 & 3 & 0 & 15 & 0 \\ 0 & 22 & 0 & 11 & 0 & 20 & 0 & 10 & 0 & 5 & 0 & 3 & 0 & 15 \\ 15 & 0 & 22 & 0 & 11 & 0 & 20 & 0 & 10 & 0 & 5 & 0 & 3 & 0 \\ 0 & 15 & 0 & 22 & 0 & 11 & 0 & 20 & 0 & 10 & 0 & 5 & 0 & 3 \\ 3 & 0 & 15 & 0 & 22 & 0 & 11 & 0 & 20 & 0 & 10 & 0 & 5 & 0 \end{bmatrix} \begin{bmatrix} f(0) \\ f(1) \\ f(2) \\ f(3) \\ f(4) \\ f(5) \\ f(6) \\ f(7) \\ f(8) \\ f(9) \\ f(10) \\ f(11) \\ f(12) \\ f(13) \end{bmatrix}$$

It is easy to check $NN^T = I \bmod 29$

We now present the data and transform block pairs for different choices of the data values.

Table 1. The 14-point NHT for modulus 29 (a=3 b=15 c=22 d=11 e=20 f=10 g=5)

| | f(n) | | g(n) | |
|---|---|---|---|---|
| 1 | 1,1,1,1,1,1,1,1,1,1,1,1,1,1 | 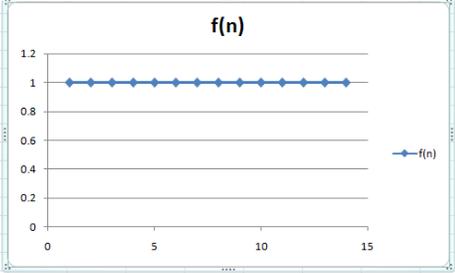 | 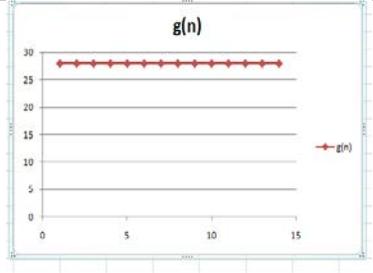 | |
| 2 | 1,1,1,1,1,1,1,0,0,0,0,0,0,0 | 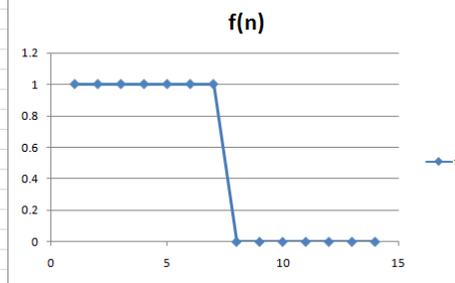 | 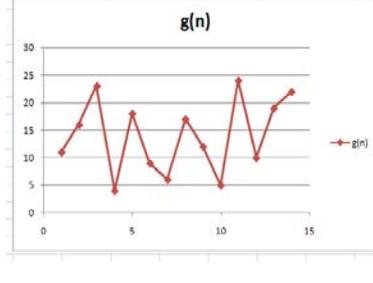 | |
| 3 | 0,0,0,1,1,1,1,1,1,1,0,0,0,0 | 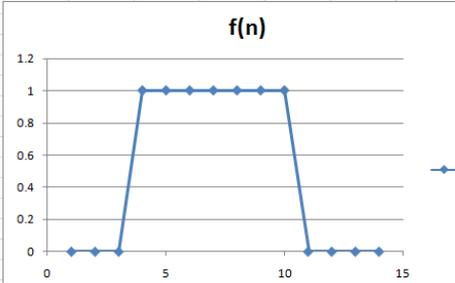 | 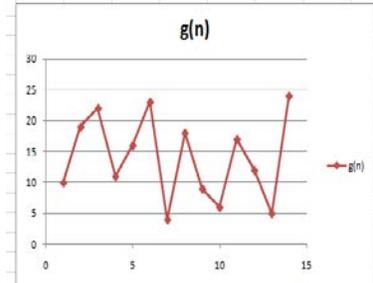 | |
| 4 | 1,1,0,0,1,1,0,0,1,1,0,0,1,0 | 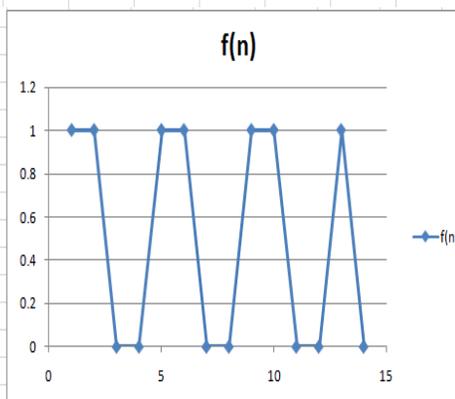 | 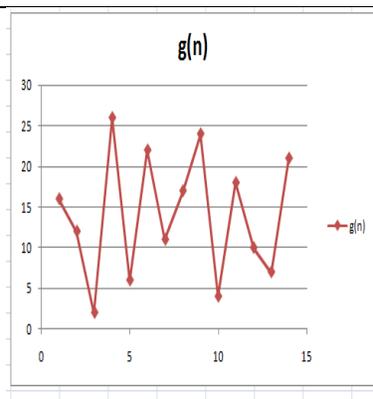 | |

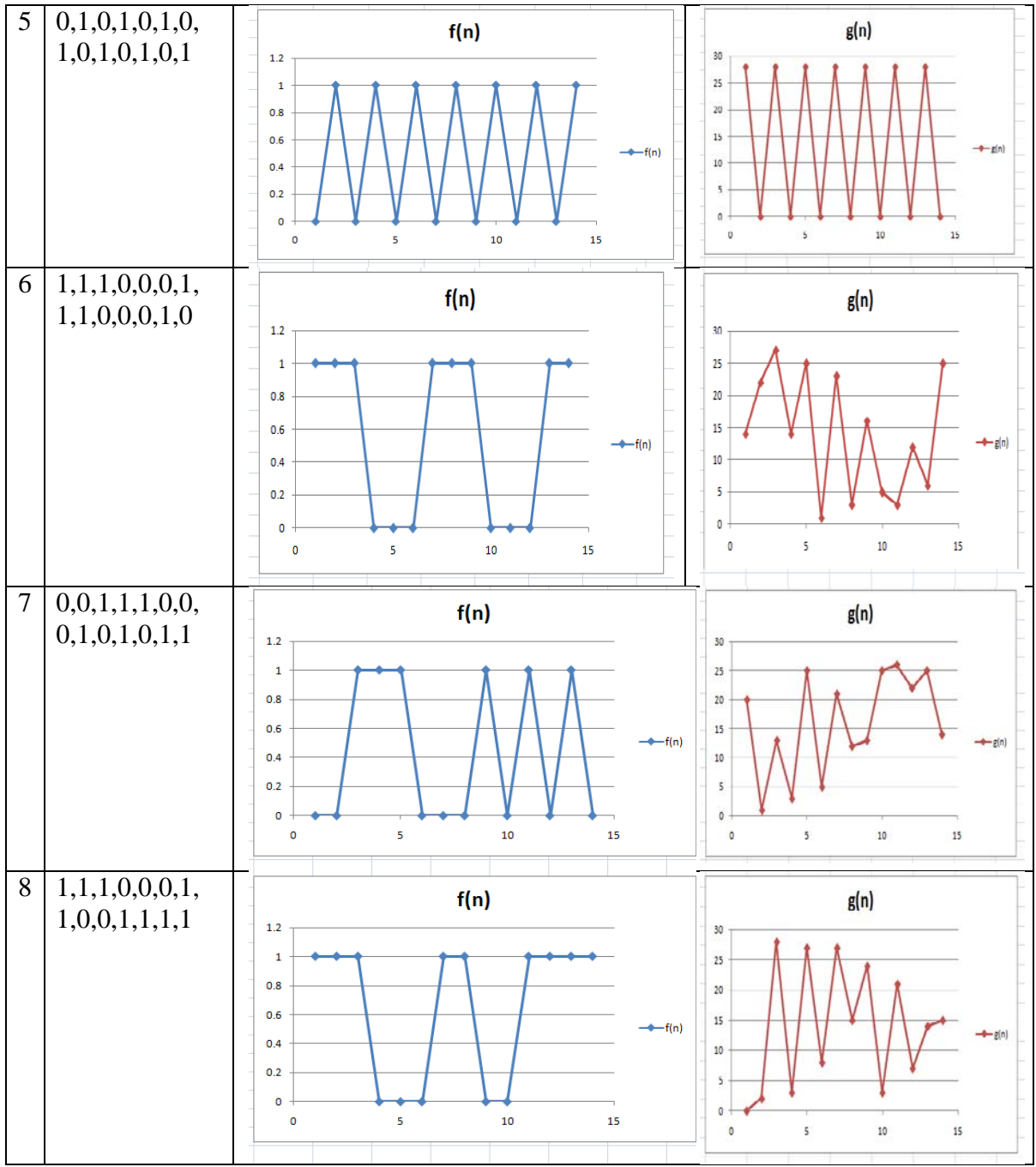

Table 2. Different solutions for the 14-point NHT

|   | a | b | C | d | e | f | g | HT mod | NN$^T$ |
|---|---|---|---|---|---|---|---|--------|--------|
| 1 | 6 | 2 | 1 | 4 | 2 | 1 | 4 | HT mod 7 | I mod 7 |
| 2 | 18 | 8 | 4 | 2 | 1 | 70 | 35 | HT mod 139 | I mod 139 |
| 3 | 134 | 110 | 63 | 126 | 95 | 33 | 66 | HT mod 157 | I mod 157 |
| 4 | 116 | 68 | 136 | 109 | 55 | 110 | 57 | HT mod 163 | I mod 163 |
| 5 | 86 | 171 | 161 | 141 | 101 | 21 | 42 | HT mod 181 | I mod 181 |

We can even assume for large numbers for values of of a =155 b =98 c =196 d =181 e =151 f=91 and f=182 and 14 point transformation will be G =HT mod 211 and can be checked as $NN^T = I \bmod 211$.

**16-point NHT**

Given the first row of the 16-point NHT matrix is 0,a,0,b,0,c,0,d,0,e,0,f,0,g,0,h and is given by the following matrix shown below. By multiplying the 16-point NHT matrix with its transpose we observe that

$$a^2+b^2+c^2+d^2+e^2+f^2+g^2+h^2$$

is the diagonal element term and the non-diagonal element terms are

(a + g) d + (e + c) h + (c + a) f + (e+ g) b,
(a+ e)(g+ c) + (h+ d)(b+ f) and
(a + g) h + (e + c) d + (c + a) b + (e+ g) f,

Table 4. NHT for different values of a b c d e f g h

|   | a  | b  | c  | d  | e  | f  | g  | h  | HT mod    | $NN^T$     |
|---|----|----|----|----|----|----|----|----|-----------|------------|
| 1 | 11 | 14 | 7  | 13 | 16 | 4  | 8  | 2  | HT mod 19 | I mod 19   |
| 2 | 34 | 67 | 45 | 1  | 2  | 4  | 8  | 16 | HT mod 89 | I mod 89   |
| 3 | 45 | 89 | 81 | 65 | 33 | 66 | 35 | 70 | HT mod 97 | I mod 97   |
| 4 | 10 | 19 | 38 | 76 | 51 | 1  | 2  | 4  | HT mod 101| I mod 101  |
| 5 | 32 | 63 | 23 | 46 | 92 | 81 | 59 | 15 | HT mod 103| I mod 103  |

Table 2. The 16-point NHT for modulus 13 (a=7 b=11 c=12 d=6 e=3 f=8 g=4 h=2)

|   | f (n)                                  |   | g(n) |
|---|----------------------------------------|---|------|
| 1 | 1,1,1,1,1,1,1,1 1,1,1,1,1,1,1,1        |   |      |

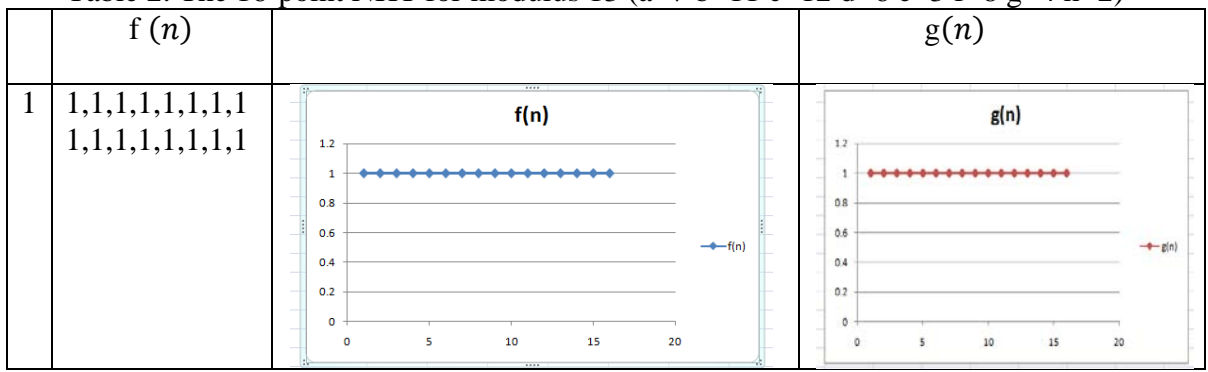

| 2 | 1,1,1,1,1,1,1,1 0,0,0,0,0,0,0,0 | 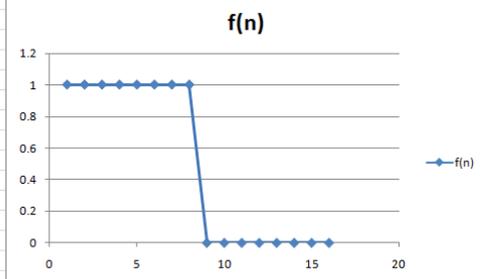 | 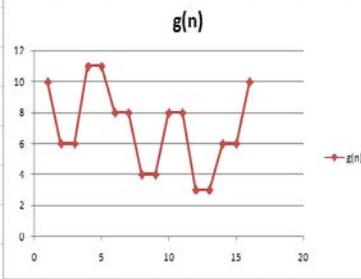 |
| --- | --- | --- | --- |
| 3 | 0,0,0,1,1,1,1,1 1,1,1,0,0,0,0,0 | 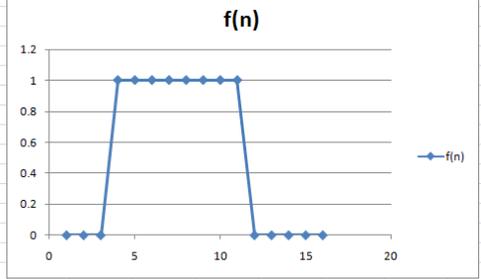 | 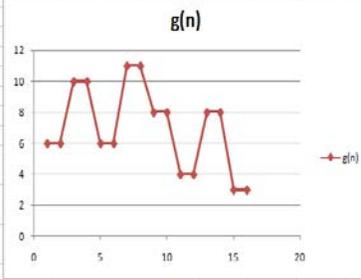 |
| 4 | 1,1,0,0,1,1,0,0 1,1,0,0,1,1,0,0 | 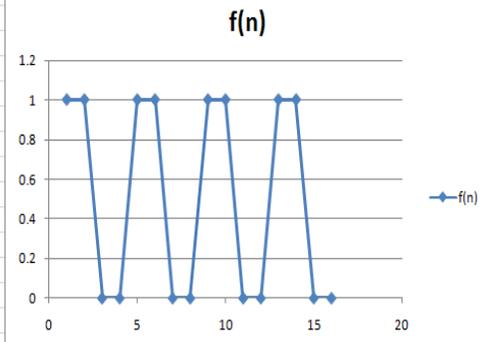 | 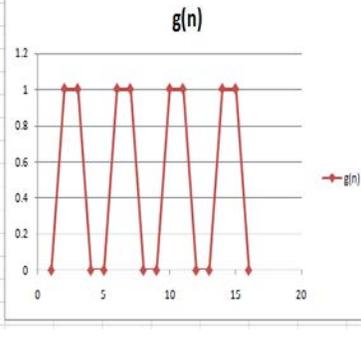 |
| 5 | 0,1,0,1,0,1,0,1 0,1,0,1,0,1,0,1 | 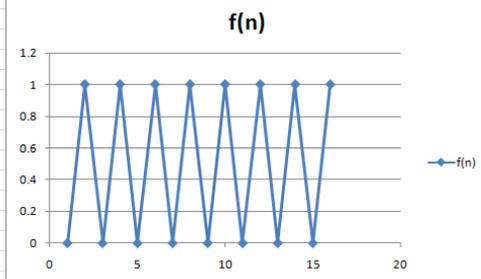 | 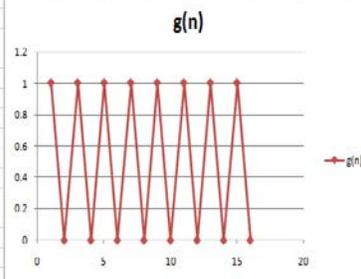 |
| 6 | 1,1,1,0,0,0,1,1 1,0,0,0,1,1,0,0 | 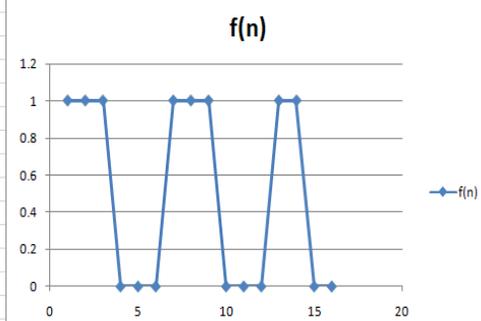 | 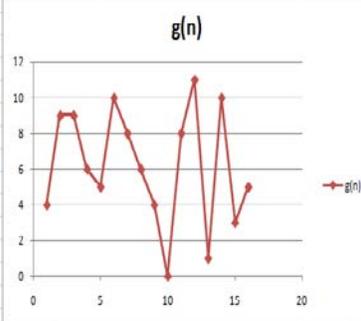 |

| 7 | 1,1,1,0,0,0,1,1 1,0,0,0,1,0,1,0 | 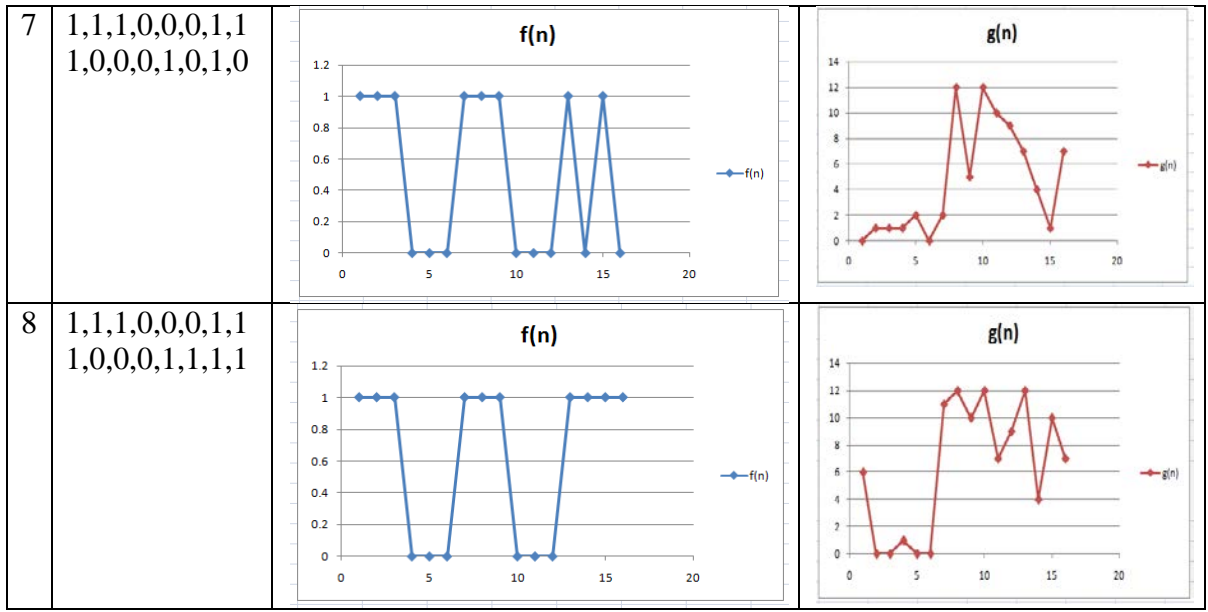 | |
|---|---|---|---|
| 8 | 1,1,1,0,0,0,1,1 1,0,0,0,1,1,1,1 | | |

To get a correct NHT matrix we need to select an appropriate matrix in such a way that all of the non diagonal elements in the $NN^T$ product matrix will become zero and only the diagonal elements will remain. There are infinite number of solutions which satisfy the equation we need to choose the values of a=7, b=11, c=12, d=6, e=3, f=8 g=4 h=2 randomly and we can write the 16-point NHT transformation as $G = HT \bmod 13$ which is as follows (with $NN^T = I \bmod 13$):

$$\begin{bmatrix} g(0) \\ g(1) \\ g(2) \\ g(3) \\ g(4) \\ g(5) \\ g(6) \\ g(7) \\ g(8) \\ g(9) \\ g(10) \\ g(11) \\ g(12) \\ g(13) \\ g(14) \\ g(15) \end{bmatrix} = \begin{bmatrix} 0 & 7 & 0 & 11 & 0 & 12 & 0 & 6 & 0 & 3 & 0 & 8 & 0 & 4 & 0 & 2 \\ 2 & 0 & 7 & 0 & 11 & 0 & 12 & 0 & 6 & 0 & 3 & 0 & 8 & 0 & 4 & 0 \\ 0 & 2 & 0 & 7 & 0 & 11 & 0 & 12 & 0 & 6 & 0 & 3 & 0 & 8 & 0 & 4 \\ 4 & 0 & 2 & 0 & 7 & 0 & 11 & 0 & 12 & 0 & 6 & 0 & 3 & 0 & 8 & 0 \\ 0 & 4 & 0 & 2 & 0 & 7 & 0 & 11 & 0 & 12 & 0 & 6 & 0 & 3 & 0 & 8 \\ 8 & 0 & 4 & 0 & 2 & 0 & 7 & 0 & 11 & 0 & 12 & 0 & 6 & 0 & 3 & 0 \\ 0 & 8 & 0 & 4 & 0 & 2 & 0 & 7 & 0 & 11 & 0 & 12 & 0 & 6 & 0 & 3 \\ 3 & 0 & 8 & 0 & 4 & 0 & 2 & 0 & 7 & 0 & 11 & 0 & 12 & 0 & 6 & 0 \\ 0 & 3 & 0 & 8 & 0 & 4 & 0 & 2 & 0 & 7 & 0 & 11 & 0 & 12 & 0 & 6 \\ 6 & 0 & 3 & 0 & 8 & 0 & 4 & 0 & 2 & 0 & 7 & 0 & 11 & 0 & 12 & 0 \\ 0 & 6 & 0 & 3 & 0 & 8 & 0 & 4 & 0 & 2 & 0 & 7 & 0 & 11 & 0 & 12 \\ 12 & 0 & 6 & 0 & 3 & 0 & 8 & 0 & 4 & 0 & 2 & 0 & 7 & 0 & 11 & 0 \\ 0 & 12 & 0 & 6 & 0 & 3 & 0 & 8 & 0 & 4 & 0 & 2 & 0 & 7 & 0 & 11 \\ 11 & 0 & 12 & 0 & 6 & 0 & 3 & 0 & 8 & 0 & 4 & 0 & 2 & 0 & 7 & 0 \\ 0 & 11 & 0 & 12 & 0 & 6 & 0 & 3 & 0 & 8 & 0 & 4 & 0 & 2 & 0 & 7 \\ 7 & 0 & 11 & 0 & 12 & 0 & 6 & 0 & 3 & 0 & 8 & 0 & 4 & 0 & 2 & 0 \end{bmatrix} \begin{bmatrix} f(0) \\ f(1) \\ f(2) \\ f(3) \\ f(4) \\ f(5) \\ f(6) \\ f(7) \\ f(8) \\ f(9) \\ f(10) \\ f(11) \\ f(12) \\ f(13) \\ f(14) \\ f(15) \end{bmatrix}$$

We can even assume for large numbers for values of of a =66 b =133 c =109 d =61 e=122 f =87 g=17 h=34 and 16 point transformation will be G =HT mod 157 and can be checked as $NN^T = I$ mod 157.

**Conclusions**

This paper presents new results in the theory of number theoretic Hilbert (NHT) transforms. New polymorphic solutions have been found for the 14-point and 16-point transforms. Several transform pairs are computed and solutions found for which the sequence and the transform have the same shape. The multiplicity of solutions for the same moduli increases their applicability to cryptography.